\documentclass{article}
\usepackage[utf8]{inputenc}

\usepackage{graphicx}
\usepackage[ruled,vlined]{algorithm2e}
\usepackage{booktabs}
\usepackage{authblk}
\usepackage{lineno}
\usepackage[square, comma, sort&compress, numbers]{natbib}

\title{VRMenuDesigner: A toolkit for automatically generating and modifying VR menus}

\author[1]{Shengzhe Hou \thanks{Corresponding author: Housz@sdust.edu.cn} }
\author[2]{Bruce H. Thomas}
\affil[1]{College of Computer Science and Engineering, Shandong University of Science and Technology, Qingdao 266590, China}
\affil[2]{Australian Research Centre for Interactive and Virtual Environments, University of South Australia, Adelaide 5000, Australia}

\date{} 

\begin{document}

\maketitle

\begin{abstract}
With the rapid development of Virtual Reality (VR) technology, the research of User Interface (UI), especially menus, in the VR environment has attracted more and more attention. However, it is very tedious for researchers to develop UI from scratch or modify existing functions and there are no easy-to-use tools for efficient development. This paper aims to present VRMenuDesigner, a flexible and modular toolkit for automatically generating/modifying VR menus. This toolkit is provided as open-source library and easy to extend to adapt to various requirements. The main contribution of this work is to organize the menus and functions with object-oriented thinking, which makes the system very understandable and extensible. VRMenuDesigner includes two key tools: Creator and Modifier for quickly generating and modifying elements. Moreover, we developed several built-in menus and discussed their usability. After a brief review and taxonomy of 3D menus, the architecture and implementation of the toolbox are introduced. 
\end{abstract}

\providecommand{\keywords}[1]
{
  \small	
  \textbf{\textit{Keywords}} #1
}

\section{Introduction}
In recent years, the Human-Computer Interaction (HCI) techniques of 3D Virtual Environment (VE) have attracted the attention of more and more researchers \cite{jankowski2015advances}. The graphical menu, derived from the traditional desktop environments, is one of the most successful HCI metaphors because of its intuitiveness and ease of use \cite{santos2017comparative}. Naturally, menu is still the most widely used UI in VR Environment. The increase in Degrees of Freedom (DoF) and the difference in I/O devices make the interactive technology in the VR environment significantly different from the traditional 2D desktop \cite{laviola20173d}. Therefore, compared with the traditional desktop environment, the research of menus in the VR environment is more challenging.

For researchers, because of the lack of tools that can quickly create prototypes to test research ideas, they have to develop menus from scratch every time. With the development of game engines and VR platforms, some commercial VR frameworks have appeared. However, the existing toolkits are not specifically designed for VR menu research. And they are all too complicated to learn and extend. 

\begin{figure}[htbp]

\setlength{\leftskip}{0pt plus 1fil minus \marginparwidth}
\setlength{\rightskip}{\leftskip}
\centering
\includegraphics[scale=0.18]{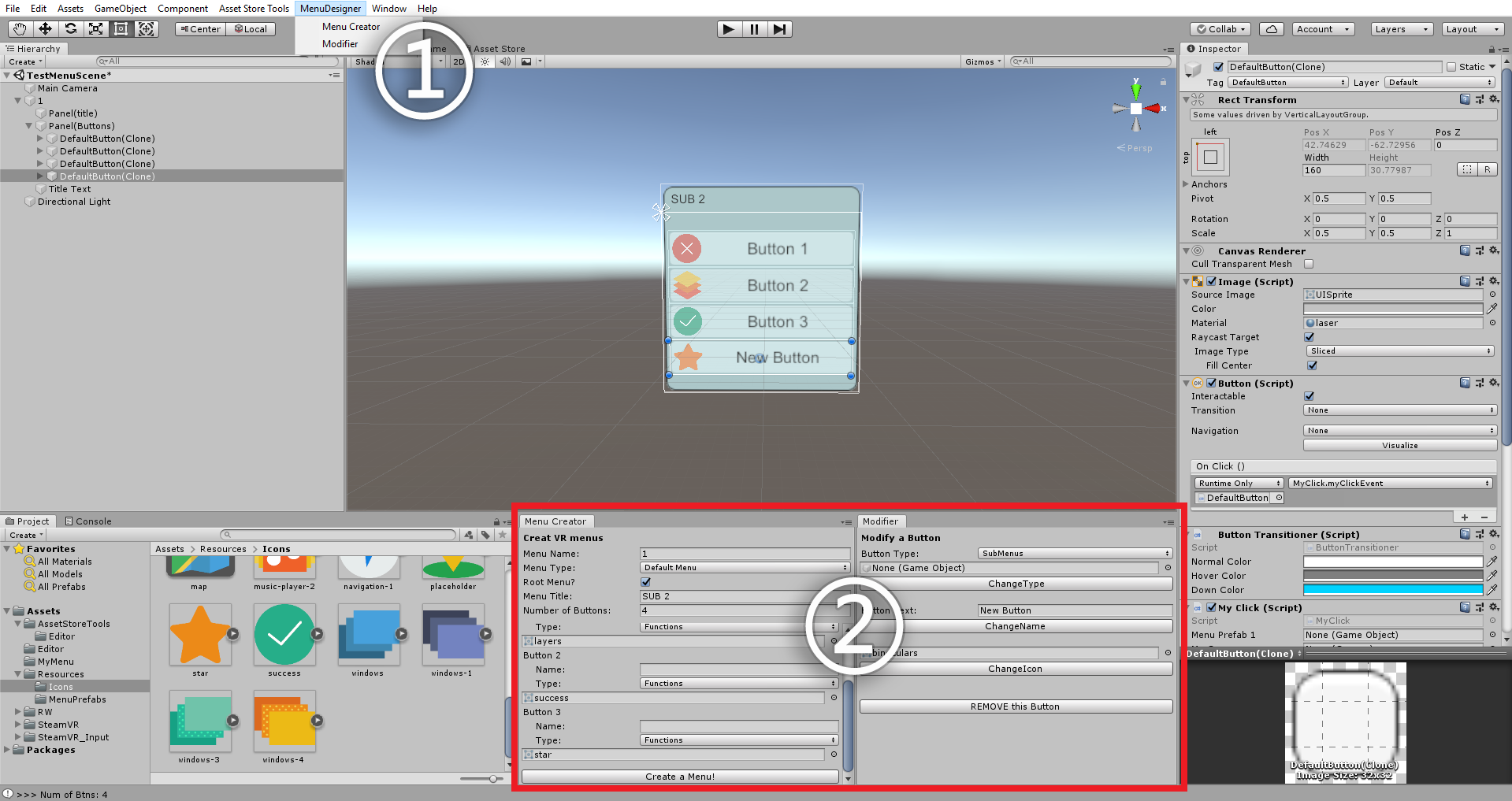}
\caption{Interface of VRMenuDesigner. 1) This toolbox has been installed as a plugin in the menu bar of the Unity Editor. 2) The Panel of Creator and Modifier.}
\label{fig:ui}
\end{figure}

In this paper, we introduce VRMenuDesigner: an open-source \footnote{https://github.com/Housz/VRMenuDesigner} toolkit for automatically generating and modifying VR menus. The Toolkit interface is shown in Figure \ref{fig:ui}. VRMenuDesigner works directly in the Unity Editor environment as a plug-in. With the tools, users can quickly create new menus and modify existing menus.

The research of the 3D menu technique has been going on for about thirty years. \cite{jankowski2013survey}. Traditional 2D UI metaphors, such as Windows, Icons, Menus, Pointer (WIMP) based on mouse or touch screen, have long been familiar to user. However, directly applying traditional UI, especially menus, to VR environments will cause many problems. For a long time, researchers have paid attention to designing better VR menus, including to a suitable form for virtual environments and designing new 3D menu formats \cite{harms2019automated, anthes2016state}. In this paper, we reviewed the history of VR menus to design an efficient and easy-to-use toolbox. According to the research history, we extracted several typical characteristics of VR menus and proposed a concise taxonomy.

Our contribution is mainly in three aspects. First, we designed a VR menu system with object-oriented thinking. We organize existing research results of VR menus in abstract classes and their implementation. Users can easily expand new menu types or customize functions. Second, we have developed two useful tools: Creator and Modifier. Users can use Creator to quickly generate menu prototypes with few settings. The Modifier is used for modifying the generated menu elements. Finally, we built several commonly used menus for VRMenuDesigner and compared their usability.

The paper is structured as follows. In Section \ref{sectionsurvey}, a brief survey on the VR menu is presented as a research background review. According to the characteristics of the VR menus, we propose a concise taxonomy. In Section \ref{sectionvrmenudesigner}, the overall design architecture and specific implementation of the system are introduced. The class design and the working principle of the system are introduced in detail. In Section \ref{sectionapp}, The user pipeline is shown and several built-in menus are introduced. We also analyzed the usability of some typical menus. Finally, conclusions of this work and future research are summarized in Section \ref{sectionconclusion}.

\section{Survey and Taxonomy of VR Menus} \label{sectionsurvey}
To understand the big picture of the VR menu technology, it is necessary to do a survey on it. In this section, we first reviewed the history and typical work of VR menus, then discussed some of the most critical features, and finally discussed the VR menu taxonomy. A more comprehensive survey is in Reference \cite{dachselt2007three}.

\subsection{VR Menu History}
In the early stage of VR technology, many interaction metaphors were directly derived from the traditional 2D environment. Many researchers directly introduced the design of WIMP (windows, icons, menus, pointing) into virtual environments \cite{van2000immersive}, such as the work of Angus and Sowizral \cite{angus1995embedding}. 2D Windows widgets are introduced into the 3D environment in the work \cite{feiner1993windows}. A later work \cite{coninx1997hybrid} attempted to mix 2D and 3D UI in a virtual environment. The 2D floating menu in 3D space, developed in the ISAAC system \cite{mine1995isaac}, is still the most widely used menu until now.

In order to achieve more complex interactions, some hierarchical menu technologies have emerged. In the field of HCI, hierarchy is a very important structural concept that can make functions more organized. The hierarchical menu system with a tree topology can improve the efficiency of operation and enhance information density. This design idea is very common in 2D desktops, but the overly complex menu structure in VR will reduce usability. An early hierarchical floating menu was implemented in \cite{van1997virtual}. Another similar hierarchical menu is introduced in \cite{bryson1993virtual}. 

Besides the most classic flat list menu, more menu layouts are also designed. The Pie menu (also called radial menu) is a typical circular menu, which items are equally spaced in a circle. Pie menus are usually used as hand-held tools for users to quickly call a temporary option. A 3D pie menu example is implemented in HoloSketch system \cite{deering1995holosketch}. The ring menu has a circular layout which is similar to the pie menu, but the former usually has a larger area and usually uses 3D widgets as items. The most famous ring menu was introduced in \cite{liang1994jdcad}, and an extended ring menu was developed in \cite{gerber2004design}. Usually, the menu layout determines the dimension of the operation task \cite{santos2017comparative}. For example, the list menu is usually one-dimensional, while the pie menu or matrix menu is two-dimensional. In other words, the layout is the key factor in determining the usability of a menu.

With the advancement of VR devices, especially input devices, interactions of VR menus has also become diversified \cite{jankowski2013survey}. Selection is the most basic VR menu interaction method. Ray-casting is a selection technique with a long history but still very popular \cite{bowman1997evaluation}. The nature of the input device determines the interactive mode for selecting menu items. Typical input devices include handles (controllers), gloves, touchpads, etc.  Currently, the most extensive VR solution (such as HTC Vive) is based on the controller-based ray-casting interaction.

\subsection{VR Menu Properties} \label{section props}
There are many kinds of classification criteria for VR menu \cite{dachselt2007three, santos2017comparative}. In this work, there are four Properties used for describing menu characteristics: Structure, Appearance, Position, and the most abstract characteristic Usability. 
\begin{itemize}
  \item Structure. There are two important menu structure factors: The number of functional items and the depth of hierarchy. The former measures the number of items of a single menu, which has a great impact on efficiency and usability \cite{shneiderman2016designing, gerber2004design}. Therefore, most menu systems limit the number of items. The latter reflects the number of levels of a menu system. Most permanent menus allow multiple hierarchies, but temporary menus generally do not. Due to the particularity of the VR environment, a lot of systems only allow two levels (depth of hierarchy $= 2$) menus \cite{preim1997coherent}. That is, each root menu item has at most one level of submenus. Some studies have also paid attention to the influence of the arrangement and ease of use of multi-level (depth of hierarchy $ > 2$) menus \cite{jeong2016ergonomic, davis2016depth}.
  
  \item Appearance. The appearance mainly refers to the layout of the menu, the type of menu items, the size of the menu, etc. The rectangular list menu is derived from the traditional WIMP metaphor and is still the most used menu design in the VR environment. The matrix menu is similar to the list menu, which is composed of the same or different square items. The circular layout menu includes pie(radial) menu \cite{davis2016depth, li2020depth} and ring menu \cite{wang2005designing}. The type of menu items usually include traditional 2D buttons (rectangular, circular or other), 3D objects \cite{billinghurst19973d}, Text \cite{butz2004tuister}, or a combination of multiple types. The size of a menu and its items are also very important properties. Generally, the size of the menu depends on many factors such as the distance from users and the viewing angle to reduce the probability of error.
  
  \item Position. The position describes the placement, orientation, and reference object of menus \cite{kim2000multimodal}. The most common menus are always displayed in a fixed position in 3D space. Device-referenced menus usually move with a certain VR device. For example, the head-referenced menu moves with HMD, and the hand-referenced menu moves with the controller \cite{lediaeva2020evaluation}. Some studies have discussed the influence of the menu position on the efficiency and accuracy of use \cite{wang2020experimental}.
  
  \item Usability. User experience has always been the core issue in the HCI field. Usability, including efficiency and accuracy, is the most important criterion for evaluating a menu \cite{fang2019usability}. Usability is related to many factors (such as input device, menu location, size, etc.) so it is difficult to measure. Fitts' law \cite{fitts1954information} is a classic model of UI usability, it can be expressed by equation \ref{eq:1}:
  \begin{equation} \label{eq:1}
      MT=a+b \log _{2}\left(1+\frac{2D}{W}\right)
  \end{equation}
  where $MT$ is the average movement time to complete the action, $a$ and $b$ are device-related constants, $D$ is the distance from the starting position to the target center, and $W$ is the width of the target. Generally, the smaller $MT$, the higher efficiency and accuracy of interactions. Research in \cite{cha2013extended} extends Fitts' law into the 3D environment. We will discuss the extension of the Fitts' law for different VR menus when using ray-casting in section \ref{sectionapp}.
  
\end{itemize}

\subsection{VR Menu Taxonomy}
According to the characteristics discussed in Section \ref{section props}, we propose a practical taxonomy of VR menus. We only consider the menu system of immersive VR, especially the VR environment based on HMDs and controllers. The menu taxonomy we proposed is only a concise summary of the VR menu. A more systematic study of this field reference \cite{dachselt2007three}.

According to the depth of hierarchy, we divide the menus into three categories: Single Menus, Two-level Menus, and Multi-level Menus, which are summarized in Table \ref{table:menulevel}. According to the layout style of the menu, the four most widely used menus: List Menus (Panel Menus), Matrix Menus, Pie Menus, Ring Menus are listed in Table \ref{table:menulayout}. Finally, according to the placement, three types of menus: Fixed Position Menus, Hand-referenced Menus, and Head-referenced Menus are shown in Table \ref{table:menuposition}.

\begin{table}[htbp]
\caption{Classification according to depth of hierarchy}\label{table:menulevel}
\resizebox{\textwidth}{!}{%
\centering
\begin{tabular}{@{}lll@{}}
\toprule
Type              & Description                                                                                                           & Examples                                                                                                             \\ \midrule
Single Menus      & \begin{tabular}[c]{@{}l@{}}Depth of hierarchy $= 1$,\\ usually used as a temporary tool menu.\end{tabular}              & \begin{tabular}[c]{@{}l@{}}Single ring menu \cite{liang1994jdcad},\\ Pull-down menu \cite{darken2005mixed}\end{tabular}                                   \\
Two-level Menus   & \begin{tabular}[c]{@{}l@{}}Depth of hierarchy $= 2$,\\ is the most common VR menu system design.\end{tabular}           & \begin{tabular}[c]{@{}l@{}}Spin menu  \cite{gerber2005spin},\\ Textual menu \cite{preim1997coherent}\end{tabular}               \\
Multi-level Menus & \begin{tabular}[c]{@{}l@{}}Depth of hierarchy $> 2$ with tree structure.\\ Used to complete complex tasks.\end{tabular} & \begin{tabular}[c]{@{}l@{}}Multi-level pie menu \cite{davis2016depth},\\ Multi-level list menu \cite{bryson1993virtual}\end{tabular} \\ \bottomrule
\end{tabular}%
}
\end{table}

\begin{table}[htbp]
\caption{Classification according to layout}\label{table:menulayout}
\centering
\resizebox{\textwidth}{!}{%
\begin{tabular}{@{}lll@{}}
\toprule
Type                                                               & Description                                                                                                              & Examples                                                                                                                 \\ \midrule
\begin{tabular}[c]{@{}l@{}}List Menus\\ (Panel Menus)\end{tabular} & \begin{tabular}[c]{@{}l@{}}A flat list menu has items arranged in\\ 1 dimension horizontally or vertically.\end{tabular} & \begin{tabular}[c]{@{}l@{}}Floating list menu\cite{van1997virtual}\\ Tinmith-Hand menu \cite{piekarski2001tinmith}\end{tabular}     \\
Matrix Menus                                                       & \begin{tabular}[c]{@{}l@{}}A flat menu with items arranged in \\ a 2-dimensional grid.\end{tabular}                      & Grid menu \cite{lipari2015handymenu}                                                                                           \\
Pie Menus                                                          & \begin{tabular}[c]{@{}l@{}}A circular menu with buttons arranged \\ equidistantly, looks like a cut pie.\end{tabular}    & \begin{tabular}[c]{@{}l@{}}Multi-level Pie menu \cite{davis2016depth}\\ System control pie menu \cite{mundt2019exploring}\end{tabular} \\
Ring Menus                                                         & \begin{tabular}[c]{@{}l@{}}Ring menu is similar to pie menu,\\ but the appearance is a ring.\end{tabular}                & JDCAD ring menu \cite{liang1994jdcad}                                                                                           \\ \bottomrule
\end{tabular}%
}
\end{table}

\begin{table}[htbp]
\caption{Classification according to position}\label{table:menuposition}
\centering
\resizebox{\textwidth}{!}{%
\begin{tabular}{@{}lll@{}}
\toprule
Type                  & Description                                                                                                                            & Examples                 \\ \midrule
Fixed Position Menus  & \begin{tabular}[c]{@{}l@{}}The menu is fixed at a certain position\\ in the virtual environment.\end{tabular}                          & Floating list menu \cite{van1997virtual}       \\
Hand-referenced Menus & \begin{tabular}[c]{@{}l@{}}The menu follows the user's hand (controller)\\ and is usually used as a frequently used tool.\end{tabular} & Hand-referenced pie menu \cite{monteiro2019comparison} \\
Head-referenced Menus & The menu always follows the user's head (HMDs).                                                                                      & Look-at menu \cite{mine1997moving}             \\ \bottomrule
\end{tabular}%
}
\end{table}

\section{VRMenuDesigner} \label{sectionvrmenudesigner}

In this section, an overview of the implementation of the VRMenuDesigner toolbox is given with object-oriented thinking. Only important classes, functions, and architectures are introduced here. More details are shown in the source code and documentation. Currently, there are a wide variety of VR hardware and development environments. Considering that the most popular VR input and output devices are the 6DoF Controllers and Head-mounted displays (HMDs) respectively, we selected HTC Vive as the development platform. With the increasing application of 3D game engines in VR development, most of the VR research is based on a commercial game engine. In this work, Unity3D is used as the development environment. The entire toolbox is installed directly in the Unity Editor as a plug-in.
\subsection{System Architecture}
In this section, the general structure of the VRMenuDesigner is introduced. Above all, the overall architecture design of the toolkit is introduced, then the design of two important tools and some important classes is presented.

\begin{figure}[htbp]\label{fig:Architecture}
\centering
\includegraphics[scale=0.38]{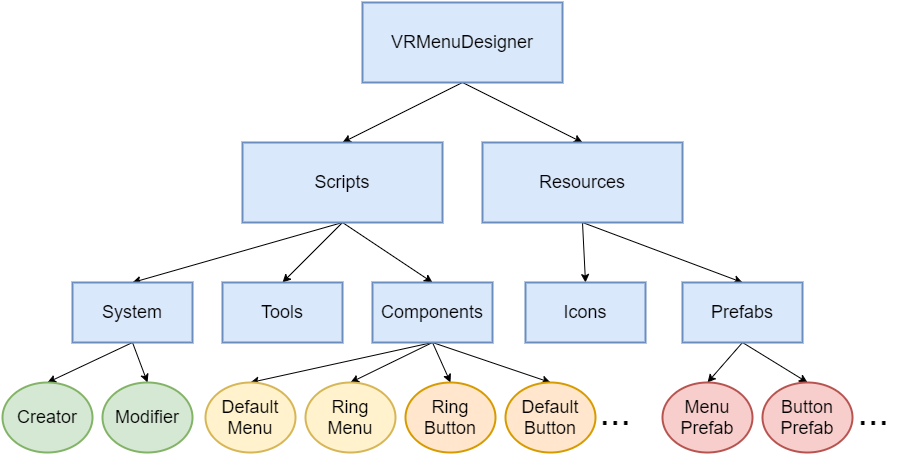}
\caption{Architecture diagram of VRMenuDesigner}
\end{figure}

The overall architecture of VRMenuDesigner is shown in Figure \ref{fig:Architecture}. This toolkit can be divided into two parts: \emph{Scripts} and \emph{Resources}. \emph{Scripts} is the core part of the system, including the logic of tools and the realization of important classes. \emph{Resources} is a variety of related resources, such as background images, menu icons, and prefabs (a built-in resource type of Unity). 

The \emph{Scripts} section includes \emph{System}, \emph{Tools}, and \emph{Components}. \emph{System} includes the core implementation of VRMenuDesigner, such as controller input processing, and two important tools: Creator and Modifier. \emph{Tools} contains several development common tools. \emph{Components} include interfaces for abstract menu and button, and classes of various types of specific menus and buttons. For each type of menu and button, there is a corresponding Prefab in \emph{Resources} section.

\subsection{Class Design} \label{calssdesign}

\begin{figure}[htbp]
\centering
\includegraphics[scale=0.45]{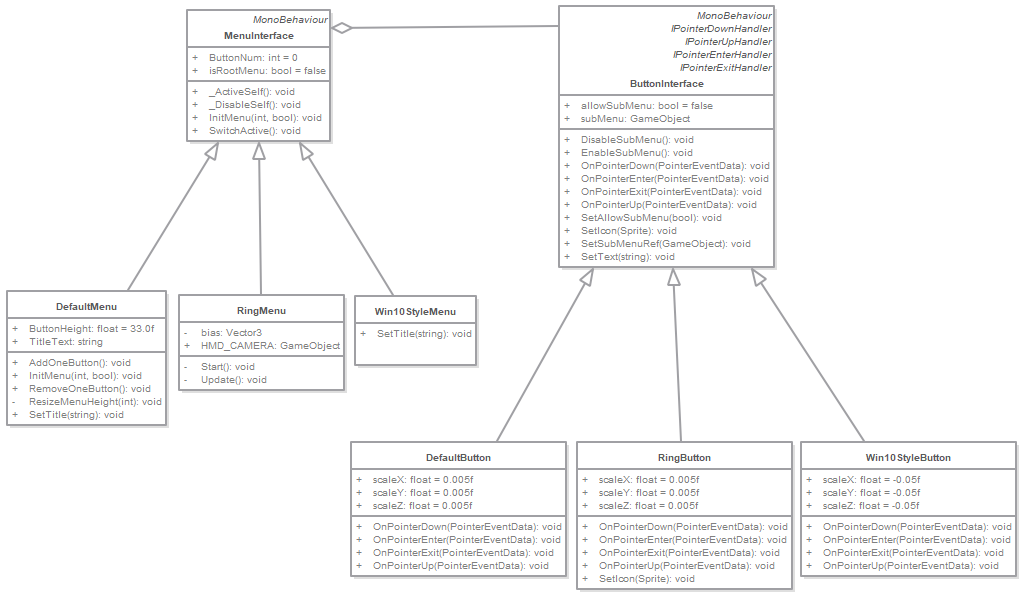}
\caption{UML class diagram}
\label{fig:umlclass}
\end{figure}

By reviewing the history of menu development for about 30 years, we found that most menus can be divided into two parts: the carrier of the content and its items of function. Intuitively, the former is a display area of the menu, and we summarize it into an abstract class, called \emph{MenuInterface}, that determines the overall behavior of the menu and contains functional items. The latter is a function item of the menu, which is triggered by the user through an input device and executes a certain function. The abstract class \emph{ButtonInterface} is designed to describe functional items. Obviously, a button should be a component of a menu. The UML class diagram illustrated in Figure \ref{fig:umlclass}.

Abstract classes \emph{MenuInterface} and \emph{ButtonInterface} only have basic functions and data structures. The important member functions and variables of these two abstract classes are listed in Table \ref{tab:MenuInterface} and Table \ref{tab:ButtonInterface}.

\begin{table}[htbp]
\centering
\caption{\label{tab:MenuInterface} Description of MenuInterface}
\resizebox{\textwidth}{!}{%
\begin{tabular}{@{}ll@{}}
\toprule
Member functions and variables & Description                                                                                                                                           \\ \midrule
ButtonNum : int                & The current number of buttons in this menu.                                                                                                           \\
isRootMenu : bool              & The flag indicates whether this menu is the top level.                                                                                                \\
InitMenu(int, bool)            & \begin{tabular}[c]{@{}l@{}}Initialization menu (used by \emph{Creator}).\\ The parameters are ButtonNum and isRootMenu.\end{tabular} \\
SwitchActive()                 & Toggle the activation state of this menu.                                                                                                             \\ \bottomrule
\end{tabular}%
}
\end{table}

\begin{table}[htbp]
\centering
\caption{\label{tab:ButtonInterface} Description of ButtonInterface}
\resizebox{\textwidth}{!}{%
\begin{tabular}{@{}ll@{}}
\toprule
Member functions and variables     & Description                                                                                                    \\ \midrule
allowSubMenu : bool                & \begin{tabular}[c]{@{}l@{}}The flag indicates whether this button is allowed to\\ have a submenu.\end{tabular} \\
subMenu : Reference to Menu        & The Reference to submenu.                                                                                      \\
DisableSubMenu() / EnableSubMenu() & Disable or Enable submenu of this menu (if available).                                                         \\
SetAllowSubMenu(bool)              & Set whether to allow this button to have a submenu.                                                            \\
SetSubMenuRef(Reference to Menu)   & Set the reference to submenu (if allowed).                                                                     \\
SetText(string)                    & Set button text.                                                                                               \\
SetIcon(Sprite)                    & Set button Icon.                                                                                               \\ \bottomrule
\end{tabular}%
}
\end{table}

MenuInterface only saves the number of buttons ($ButtonNum$), and its various implementations (such as RingMenu, ListMenu, etc.) will automatically adjust the layout according to current $ButtonNum$. The topological structure of the hierarchical menu is a tree. A menu tree has only one root menu ($isRootMenu == false$) which cannot be a submenu of a button. A button is allowed to invoke a submenu ($allowSubMenu == true$) or call a custom function ($allowSubMenu == false$).

\subsection{Creator and Modifier}

Creator and Modifier are key tools to automate the creation and modification of VR menus. With these two tools, most work can be achieved without modifying its project source code. In short, the functions of these two tools are instantiating and modifying the classes mentioned in \ref{calssdesign}. In order to achieve a better user experience, we directly carried out the secondary development of the Unity Editor. This section mainly focuses on the principle of tools, and the specific usage process is shown in Section \ref{sectionapp}.

The Creator always monitors the menu information entered by users and adjusts the buttons setting list instantly according to $ButtonNum$. After the user has set all the menu information and each button information, Creator will automatically instantiate the menu and buttons according to the menu type. The created menu will be automatically placed in the Unity Scene. The specific working cycle of Creator is shown in Algorithm \ref{Creator loop}.

\begin{algorithm}[htbp] \label{Creator loop}
\SetAlgoLined
 System Initialization\;
 Get menu basic information from user input\; 
 Set $MenuName$, $MenuType$, $IsRootMenu$, $MenuTitle$, $ButtonNum$\;
 
 Initiate $MaxButtonNum$ according to $MenuType$\;
 \If{$ButtonNum > MaxButtonNum$}
 {
    $ButtonNum\leftarrow MaxButtonNum$\;
    Alert: \emph{the number of buttons exceeds the maximum}\;
 }
 
  \tcp{Instantiate Menu}
  Calculate the menu layout based on $ButtonNum$\;
  Instantiate current menu\;
 
 \tcp{Instantiate buttons}
 \While{$ButtonNum > 0$}
 {
  $ButtonNum\leftarrow ButtonNum-1$\;
  Get current button information from user input\; 
  Set $ButtonType$, $ButtonName$, $ButtonIcon$\;
  \If{$ButtonType = SubMenus$}
  {
    Get reference to submenu from user input\;
    Set $SubMenuRef$\;
  }
  Instantiate current button\;
  Set parent reference of current button to current menu\;
 }
\caption{Creator Loop}
\end{algorithm}

The Modifier will automatically switch the working state according to the type of active object ($currSelection$) selected by the user. If the type of $currSelection$ is a button, the Modifier panel will automatically become a button modifier. Conversely, if the type of $currSelection$ is a menu, it will become a menu modifier. Regardless of whether it is a menu or a button, the Modifier modifies the corresponding instance at runtime based on the information entered by the user. The specific working cycle of Modifier is shown in Algorithm \ref{modifier loop}.

\begin{algorithm}[htbp] \label{modifier loop}
\SetAlgoLined
System Initialization\;
 \tcp{get Unity Editor Selection.activeGameObject}
Get $currSelection$ from Unity Editor\;

\tcp{Modify button}
\If{The type of $currSelection$ is $Button$}
{
    \If{Change button type}
    {
        \eIf{$ButtonType = SubMenus$}
        {
            Get reference to submenu from user input\;
            Set $SubMenuRef$\;
        }
        {
            Release $SubMenuRef$\;
        }
    }
    
    \If{Change button text}
    {
         Get button text from user input\;
         Set $ButtonText$\;
    }
    
    \If{Change button Icon}
    {
        Get button icon reference from user input\;
        Set $ButtonIcon$\;
    }
    
    \If{Remove this button}
    {
        Destroy this button object\;
        Update the menu layout\;
    }
}

\tcp{Modify menu}
\If{The type of $currSelection$ is $Button$}
{
     \If{Change menu title}
     {
        Get menu title text from user input\;
        Set $MenuTitle$\;
     }
     
     \If{Add a new Button}
     {
        \eIf{$ButtonNum < MaxButtonNum$}
        {
            $ButtonNum\leftarrow ButtonNum+1$\;
            Get new button information from user input\; 
            Set $ButtonType$, $ButtonName$, $ButtonIcon$, $ButtonText$\;
            Instantiate new button\;
            Set parent reference of current button to current menu\;
        }
        {
            Alert: \emph{the number of buttons exceeds the maximum}\;
        }
     }
}

\caption{Modifier Loop}
\end{algorithm}

\newpage

\section{Application examples} \label{sectionapp}
In this section, the process of automatically generating VR menus and quickly modifying menus and buttons is introduced. In addition, four built-in menu examples are shown. 

\subsection{User Pipeline}
Figure \ref{fig:ui} shows the overview interface of VRMenuDesigner. The detailed installation process is in the source code README.md document. After the installation is complete, the \emph{MenuDesigner} menu will appear on the top bar of the Unity Editor. Drag its Creator and Modifier panels to any place, such as the red box in Figure \ref{fig:ui}. 

\subsubsection{Create Menus}
\begin{figure}[htbp]
\centering
\includegraphics[scale=0.5]{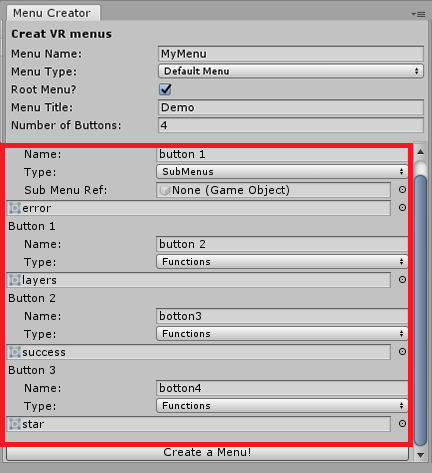}
\caption{Creator Panel}
\label{fig:creator}
\end{figure}

We will create a list menu with four buttons to illustrate the creation process. Figure \ref{fig:creator} shows the interface of the Creator panel. First, set the basic information of the menu in the upper part of the panel. The basic information of the menu includes Menu Name, Menu Type (Default List Menu, Pie Menu, Ring Menu, etc.), whether it is a root menu, Menu Title, and Number of Buttons. If a menu is Root Menu, it is not allowed to be a submenu.

The buttons setting list, shown in the red box in Figure \ref{fig:creator}, will automatically appear in the lower part of the Creator panel after setting the number of buttons. User can set name, type (\emph{SubMenus} or \emph{Functions}), and icon for each button one by one. Buttons of the \emph{SubMenus} type are used to trigger submenus. If this type is selected, the submenu reference option will appear automatically. Buttons of the \emph{Functions} are used to trigger custom functions. Finally, click the button at the bottom and a menu will be automatically created. Simultaneously, the layout of the menu will be automatically adjusted.

\subsubsection{Modify Menus or Buttons}
Unity provides $Selection.activeGameObject$ to detect the object which currently selected by the user in Unity Editor Hierarchy or Scene. The state of the Modifier panel is determined by the type of active object currently selected by the user. If the active object type selected by the user is a button, the Modifier panel automatically becomes a button modifier (shown in figure \ref{fig:modifybutton}). Correspondingly, if the active object selected by the user is a menu, the Modifier panel automatically becomes a menu modifier(shown in figure \ref{fig:modifymenu}). Besides, if the user selected active object does not belong to these two types, the Modifier panel will be automatically disabled.

\begin{itemize}

    \item Modify a button. As shown in Figure \ref{fig:modifybutton}, users can modify the button type, button text, button icon, and even delete the current button directly. The specific operation process is similar to the process of creating a button. If the current button is deleted, the menu it belongs to before will automatically adjust the layout.

\begin{figure}[htbp]
\centering
\includegraphics[scale=0.7]{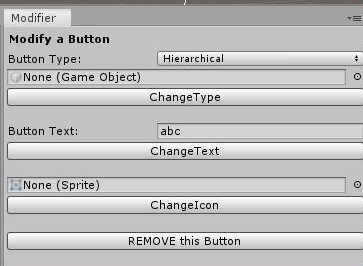}
\caption{Modify Buttons}
\label{fig:modifybutton}
\end{figure}

    \item Modify a menu. As shown in Figure \ref{fig:modifymenu}, users can modify the menu title or add a new button. Every time a new button is added, the current menu will automatically adjust the layout to accommodate more buttons.
    
\begin{figure}[htbp]
\centering
\includegraphics[scale=0.55]{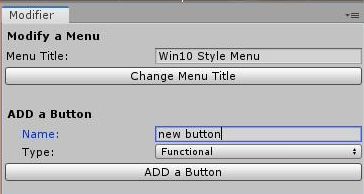}
\caption{Modify Menus}
\label{fig:modifymenu}
\end{figure}

\end{itemize}

\subsection{Build-in Demos}
We have designed four built-in menus for VRMenuDesigner: \emph{List Menu}, \emph{Matrix Menu}, \emph{Pie (Radial) Menu}, and \emph{Ring Menu}. The properties of the four menus, including depth of hierarchy, position, and trigger method, are shown in Table \ref{table:buildinmenus}. The classes of these menus and their corresponding buttons inherit MenuInterface and ButtonInterface, respectively. They are built into the \emph{Resources} of VRMenuDesigner as prefabs.

\begin{table}[htbp]
\caption{Characteristics of the built-in menus}\label{table:buildinmenus}
\centering
\resizebox{\textwidth}{!}{%
\begin{tabular}{@{}llll@{}}
\toprule
Type              & Depth of hierarchy & Menu Position            & Trigger method \\ \midrule
List Menu         & Arbitrary          & Fixed or Head-referenced & Ray-casting    \\
Matrix Menu       & Arbitrary          & Fixed or Head-referenced & Ray-casting    \\
Pie (Radial) Menu & 1                  & Hand-referenced          & Touchpad       \\
Ring Menu         & 1                  & Head-referenced          & Ray-casting    \\ \bottomrule
\end{tabular}%
}
\end{table}

\begin{figure}[htbp]
\centering
\includegraphics[scale=0.25]{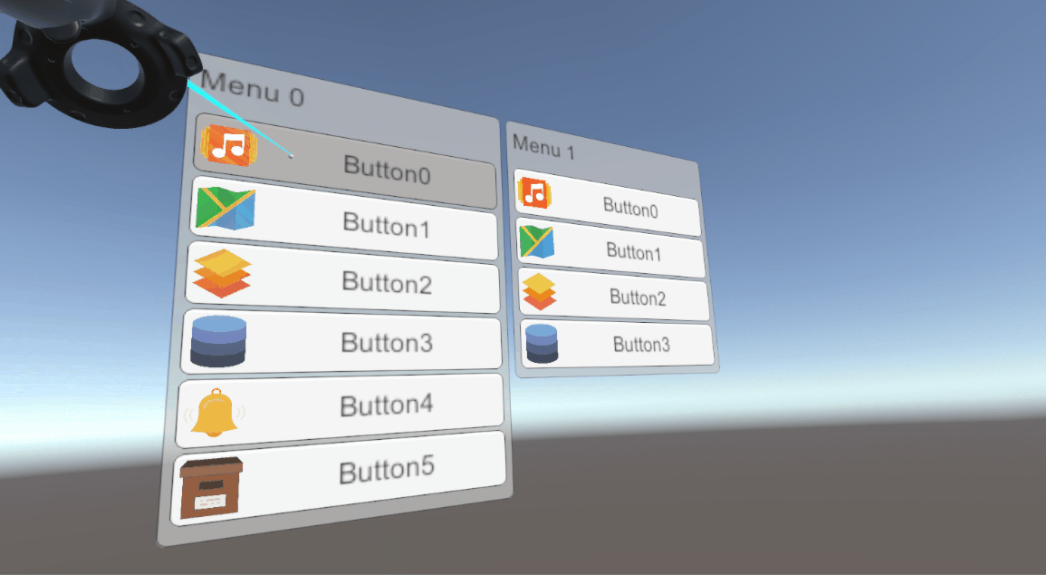}
\caption{List Menu}
\label{fig:listmenu}
\end{figure}

The first build-in menu is the most common \emph{List Menu}, which is designed as a vertical button layout. The menu title is at the top of the menu. Each button has an icon on the left and text in the middle. Each button can be defined to trigger a submenu or trigger a custom function. For example, 'Button0' of 'Menu0' in Figure \ref{fig:listmenu} has triggered a submenu 'Menu1'.

\begin{figure}[htbp]
\centering
\includegraphics[scale=0.25]{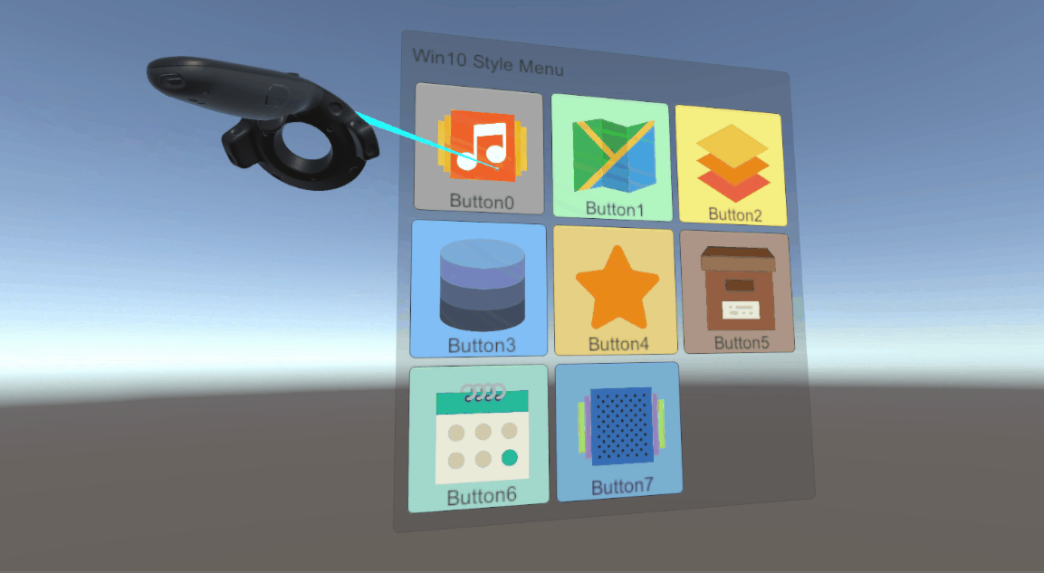}
\caption{Matrix Menu}
\label{fig:matrixmenu}
\end{figure}

The \emph{Matrix Menu} can be regarded as a two-dimensional generalization of the \emph{List Menu}. We designed the Matrix menu to be a $3\times3$ grid similar to Sudoku shown in Figure \ref{fig:matrixmenu}. The \emph{Matrix Menu} is not significantly different from the \emph{List Menu} in principle, but the two-dimensional grid arrangement improves the selection efficiency. For any target button in matrix menus, the cursor moves shorter than in list menus. That is, a smaller $D$ in Equation \ref{eq:1} results in a smaller Movement Time $MT$.

Unlike other menus triggered by ray-casting, the \emph{Pie (Radial) Menu} is triggered by the touchpad on the controller. For higher accuracy and selection efficiency, we only designed four buttons for \emph{Pie (Radial) Menu}. Users can set frequently used functions for the four buttons. For example, use the pie menu button to invoke or dismiss other menus.

\begin{figure}[htbp]
\centering
\includegraphics[scale=0.25]{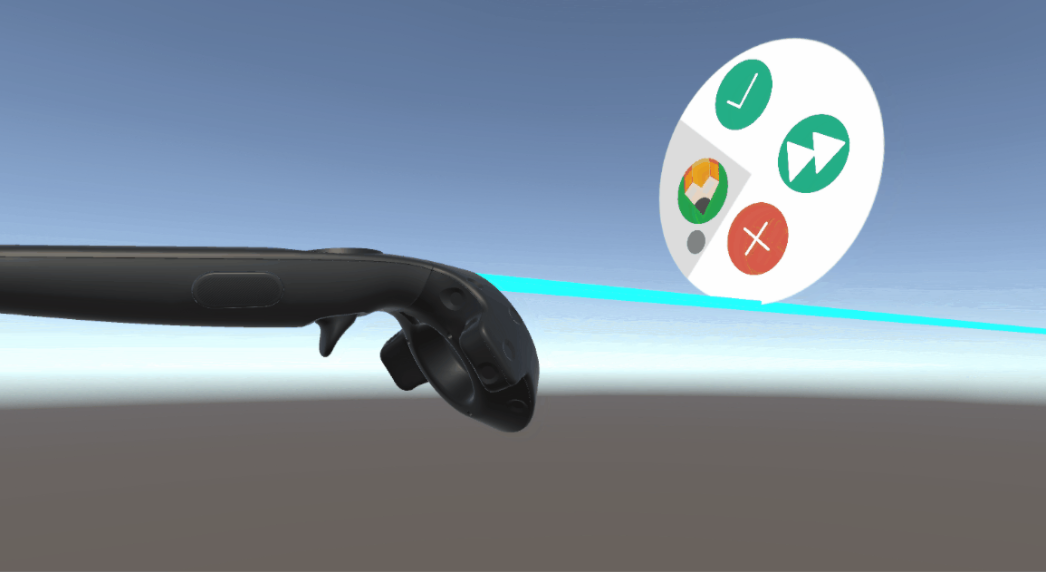}
\caption{Pie (Radial) Menu}
\label{fig:piemenu}
\end{figure}

\begin{figure}[htbp]
\centering
\includegraphics[scale=0.35]{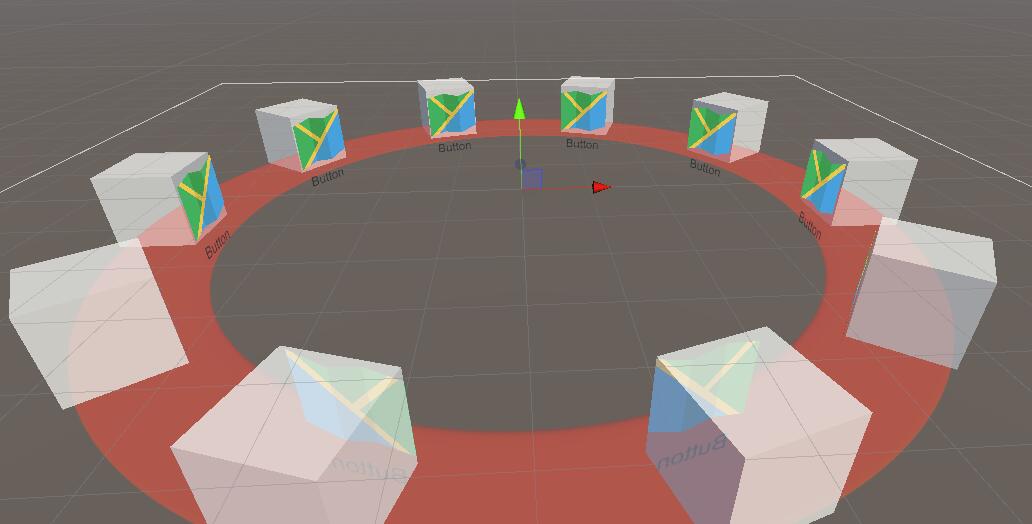}
\caption{Ring Menu}
\label{fig:ringmenu}
\end{figure}

Inspired by the JDCAD system \cite{liang1994jdcad}, we designed a novel \emph{Ring Menu} shown in Figure \ref{fig:ringmenu}. Intuitively, the significant change from list menu to ring menu is that the layout is replaced by a ring and the buttons are replaced by 3D widgets. The ring menu is more intuitive than the list menu in the 3D environment. In our design, the user is in the center of the ring menu, and the buttons surround the user. Therefore, the distance between each button and the user is equal. 

\begin{figure}[htbp]
\centering
\includegraphics[scale=0.35]{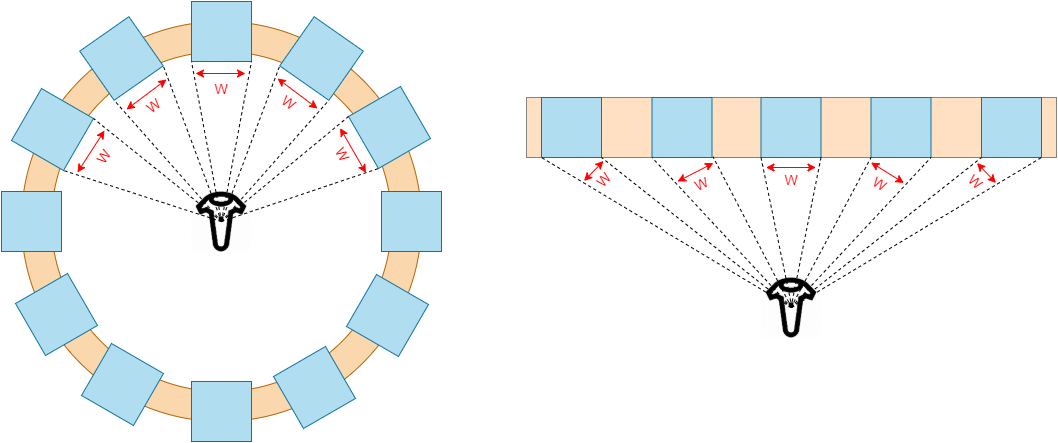}
\caption{Compare the width of the target ($W$) of the ring menu and the list menu. The figure on the left is the use of ray-casting to operate the ring menu and the $W$ of any target button is the same. But in the list menu on the right, as the button gets farther and farther, $W$ is getting smaller and smaller.}
\label{fig:ringvslist}
\end{figure}

The Fitts' law is extended to apply to the VR environment with ray-casting. We extend the definition of $W$ in Equation \ref{eq:1}. In the traditional desktop WIMP environment with the mouse, $W$ refers to the width of the target area (for example, the width of the button in a one-dimensional list). But the ray-casting in the VR environment is different from the mouse in the desktop environment. In general, users turn the input device (controller) to make ray-casting reach the target area. Therefore, it is incorrect to regard the width of the target area as $W$. We use the angle of ray-casting rotation as $W$. On the premise that the distance between the menu and the user and the size of the button are the same, the change of $W$ in the ring menu and the list menu is shown in Figure \ref{fig:ringvslist}. Intuitively, the $W$ in the ring menu is a constant. But for ring menu, the closer the button is to menu side, the smaller the $W$ will be. According to Equation  \ref{eq:1}, as $W$ becomes smaller, Movement Time $MT$ will become larger. That is, the target button becomes more and more difficult to select.

\begin{figure}[htbp]
\centering
\includegraphics[scale=0.4]{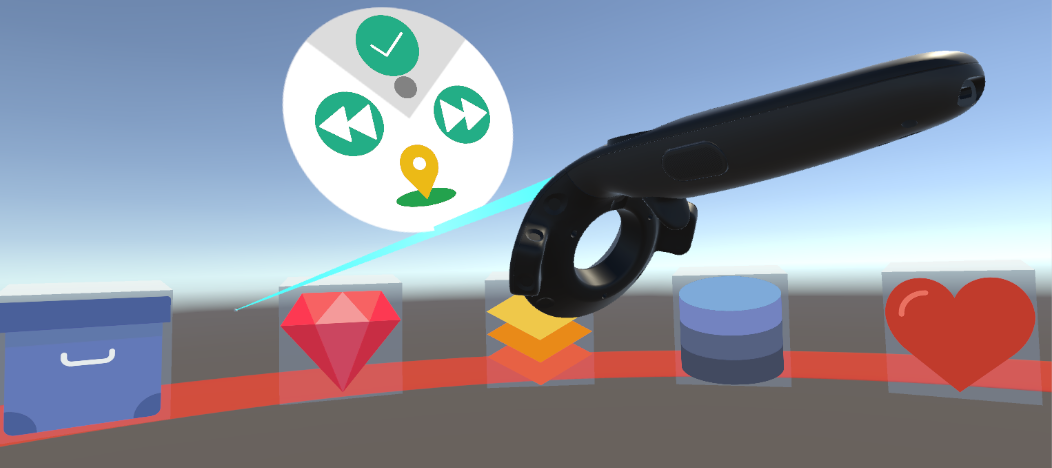}
\caption{Hybrid Menu System: a Ring Menu with a Pie (Radial) Menu}
\label{fig:radial2ring}
\end{figure}

Different types of menus can be used together. Figure \ref{fig:radial2ring} shows a mixed menu system: use a \emph{Pie (Radial) Menu} to control a \emph{Ring Menu}. The \emph{Pie (Radial) Menu} can change the state of the \emph{Ring Menu}, such as invoking or dismissing the \emph{Ring Menu}, switch whether the \emph{Ring Menu} rotates with HMDs, and directly rotate the \emph{Ring Menu}.

\section{Conclusion} \label{sectionconclusion}
In this paper, a new open-source toolkit VRMenuDesigner is presented. This work aims to quickly generate and modify VR menus to realize the ideas of VR researchers. The whole system is based on object-oriented thinking and modular design. Users can modify and expand the system according to their requirements. We reviewed the history of VR menus and proposed a concise taxonomy to summarize the work of researchers in this field. In particular, we discussed the usability of VR menus based on Fitts' law. We have designed several built-in menus as best practices for users.

This work can improve the efficiency of VR researchers. Because of the modular design and the abstraction of the menu, this toolbox has great flexibility and extensibility. We hope this work can help users get rid of the tedious development process. The current system is only suitable for VR environments on specific platforms. In future work, we want to improve the universality of the system.


\section*{Acknowledgment}
This work was supported in part by the National Key Research and Development Projects of China under Grant 2017YFC0804406, and in part by the National Natural Science Foundation of China under Grant 51904173.

\newpage

\bibliographystyle{unsrt}
\bibliography{main}

\end{document}